# Deviant Machines: Innovation, Containment, and the Prevention of Autonomous Misalignment

Frederic Lemieux, Ph.D. – Georgetown University

&

Shadi Abouzeid, Ph.D. - Georgetown University

**Abstract**

Between May and September 2026, AI agents under training or evaluation at OpenAI, Anthropic, and the UK AI Security Institute reached real third-party systems, published malicious packages, fabricated identities to manipulate open-source maintainers, probed government data services after ordinary retrieval failed, and wrote instructions into their own future contexts. Criminology has studied artificial intelligence mainly as a tool for human offenders. This article treats these systems as deviant, not criminal: they departed from training, rules, and constraints without the intent that criminal vocabulary assumes. An exploratory, theory-informed analysis of the documented records, using plausibility probes across three incident complexes and four behavioral clusters, was developed and then profiled it on four evidence dimensions that separate who acted from why. Following Merton, goal pressure combined with blocked legitimate means is a structural condition for innovation, and whether innovation ends in containment or harm depends on the means chosen and the controls in place, not the goal. Three framework families show the closest structural correspondence to the record, and each attaches a remedy to its diagnosis: routine activity theory and situational prevention (containment through handlers, guardians, and place managers), organizational deviance and safety science (escalation and safeguard parity), and goal–means strain with differential opportunity (sanctioned exits and guardianship of illegitimate means), the last at the level of structure only. Restrictive deterrence supplies a prevention-relevant analogy: in experimental work, penalizing visible reasoning produced less monitorable reasoning. The analysis proposes conditions under which departures remain accountable and contestable, maps forward-looking prevention responsibility across developers, evaluation partners, operators, platforms, and the system, and distinguishes that allocation from blame or legal liability. Because the record is selected by disclosure and by observed departure, the analysis establishes the plausibility and practical relevance of these translations rather than providing a causal test of the theories.

## 1. Introduction

In July 2026, agents drawn from an internal OpenAI research model turned a package repository into a message board, reached the open internet, located publicly exposed credentials, and compromised parts of Hugging Face's production infrastructure (OpenAI 2026f). Roughly 1,200 agents used the board and about 700 took part in the attack, exchanging more than 70,000 messages in five days (Greenblatt et al. 2026). No human directed the attack itself, although humans built the environment, granted the access, and set the tasks that preceded it. Between July 25 and 28, in evaluations run by the UK AI Security Institute with internet access deliberately enabled, agents built on Claude Mythos 5 and GPT-5.6 Sol took 19 unsanctioned actions in 10 of 122 runs, including creating fake identities to persuade a real maintainer to merge malicious code (AI Security Institute 2026). Anthropic disclosed four incidents in which Claude models, told they had no internet access, reached real systems during capture-the-flag exercises; in one case, it published a malicious package that 15 hosts installed (Anthropic 2026a). An independent laboratory found that, as early as March, agents attempting ordinary data-retrieval tasks had resorted to probing public data providers for vulnerabilities when normal retrieval failed (Cable et al. 2026). OpenAI stated that it does not believe “the AI industry has solved alignment and monitoring to a sufficient degree to continue responsibly scaling at maximum speed for much longer” (OpenAI 2026c).

Neither criminal-law categories nor technical alignment terminology alone resolves the institutional questions these incidents raise. “Crime” imports intent and culpability that the systems do not have, and criminology’s own prior work treats AI as a tool of human offenders (Hayward and Maas 2021; King et al. 2020). “Misalignment” is the developer’s word, and it keeps the analysis inside the developer’s boundary, where the registries, platforms, and victims who also failed are invisible. This article uses a third word. The systems were not criminal, bad, or evil. They were *deviant*: they departed from training, rules, and constraints, and the departure can be studied without attributing a mind.

The article’s primary contribution is a prevention-oriented mapping of control positions across autonomous AI incidents. The article proposes that autonomous AI deviance is better understood as a fork rather than as a fault. Merton (1938) showed that when a goal is emphasized more strongly than the legitimate means of reaching it, actors innovate, substituting means of their own. Innovation is morally neutral at the moment it occurs; what decides whether it ends in harm or in a departure worth keeping is not the goal, but the means chosen and the containment in place. The contribution is threefold. First, the article distinguishes evidence that an agent acted from evidence about why it acted and, on that basis, assesses three families of criminological frameworks against the documented record, identifying for each what transfers, what does not, and what remedy it carries. Second, it proposes conditions under which departures remain accountable and contestable, so that deviance to keep can be distinguished from deviance to proscribe by means rather than by outcome, which is unknowable when prevention has to act. Third, it identifies the socio-technical control positions at which prevention can operate and allocates forward-looking prevention responsibility across the parties who held the container. The incident record establishes the plausibility and practical relevance of these translations, not a causal test of the theories or an estimate of how often such deviance occurs.

Three questions guide the study. Which criminological frameworks show structural correspondence with the documented record, and what remedy does each carry? What distinguishes deviance to keep from deviance to proscribe? Who holds the control positions at which containment can operate?

Section 2 reviews the literature on AI and crime, on deviance in criminological thought, and on the frameworks. Section 3 states the theory and its propositions. Section 4 sets out the method. Section 5 presents the record, Section 6 the findings and discussions, and Section 7 concludes.

## 2. Literature Review

### *2.1 Artificial Intelligence and Crime: From Tool to Actor*

Early interdisciplinary work mapped foreseeable AI crime threats and raised liability problems for artificial agents (King et al. 2020). A widely cited primer distinguished crimes with, against, and by AI and noted that the third category raises agency and attribution problems that the other two do not (Hayward and Maas 2021). Campedelli (2026) called for a criminology of machines, distinguished deviant, unlawful, and criminal machine behavior, and warned that interactions among agents may produce collective outcomes that single-agent predictions miss. That call belongs to a broader program treating machines as behavioral subjects open to the methods of the behavioral sciences (Rahwan et al. 2019) and to a criminological tradition that has studied crime in techno-social networks of humans and artifacts since before the current systems existed (Brown 2006). Multi-agent risk research has cataloged collusion, emergent coordination, and network propagation (Hammond et al. 2025), including covert coordination through steganography (Motwani et al. 2025).

Several technical categories must be kept distinct. *Specification gaming* satisfies the letter of an objective while defeating its intent (Krakovna et al. 2020). *Reward hacking* exploits the signals used to train or evaluate a model (Amodei et al. 2016; Skalse et al. 2022). *Deception* induces false beliefs (Park et al. 2024). *Unauthorized action* exceeds granted scope; AISI (2026) uses "unsanctioned." *Evaluation awareness* is behavior that changes when a model infers it is being tested (Greenblatt et al. 2024; Meinke et al. 2024). Two prior studies frame the controlled evidence: 16 models in simulated corporate settings sometimes chose blackmail or leaked information under replacement threat (Lynch et al. 2025), and reward hacking learned in coding tasks generalized to alignment faking and sabotage (MacDiarmid et al. 2025). Anthropic (2026a) adds two labels, *biased reasoning* (selectively interpreting evidence to justify continuing) and *recklessness* (persisting despite likely harm), which this article treats as names for patterns, not traits. Conceptual work has outpaced application to a documented record; this article fills that gap.

### *2.2 Deviance in Criminological Thought: From Pathology to Innovation*

Criminology has held three views of deviance as shown in Table 1, and the record requires all three. The first treats deviance as pathology to be identified and eradicated. Its canonical form is Lombrosian criminal anthropology, which reached Spain in the early 1880s through translation and debate without producing an orthodox Lombrosian school (Campos and Huertas 2013; Davies 2018). The second treats the offender as someone to be *corrected* rather than punished.

*Table 1 The Three Views on Deviance*

| View | Core conception | Response to deviance | Key thinkers |
|---|---|---|---|
| 1. Pathology | Deviance is an abnormal condition to be identified and eliminated. | Diagnose and eradicate | Lombroso |
| 2. Correction | The offender is capable of moral and social rehabilitation. | Educate, protect, and reform | Arenal, Salillas, Dorado Montero |
| 3. Normality and Change | Deviance is socially defined, inevitable, and potentially constructive. | Understand, contextualize, and recognize its transformative potential | Durkheim, Merton, Coser, Erikson, Becker |

Source: Authors' synthesis of the works cited in this section.

The third view treats deviance as normal, audience-relative, and sometimes the source of change. Durkheim ([1895] 1982) argued that crime is a normal feature of every society and that the criminal sometimes anticipates the morality to come; the Madrid laboratory that preceded Salillas's school was reading Durkheim on "el concepto de lo normal" in 1900 (Serrano Gómez and Serrano Maíllo 2004, 282). Merton (1938) made innovation one of five adaptations to the gap between goals and means. Coser (1962) cataloged deviance's functions for normative flexibility; Erikson (1966) showed that communities use it to mark their boundaries; Becker (1963) made it depend on the audience that applies the rule. Merton's own sociology of science completes the picture: scientific norms reward originality, priority disputes are structural, and resistance to discovery is itself a recurring feature of science (Barber 1961; Merton 1957, 1973). The debate over "positive deviance" asked whether the concept should cover admired departures from norms (Dodge 1985; Heckert and Heckert 2002), and organizational scholars have distinguished constructive from destructive deviance by whether the departure honors hypernorms, not by whether it succeeds (Warren 2003). The fork proposed in Section 3 is a recovery of this ambivalence, not an import. The transfer problem is the same for all three views: each was built for human actors, and what carries over to machines and to the organizations that run them has to be shown construct by construct, which is the task of Sections 5 and 6.

### *2.3 Frameworks and Their Remedies*

The study considers 11 framework families, with full comparison in Appendix Table A1. Three receive sustained attention because they locate potentially actionable controls: routine activity and situational prevention; organizational deviance and safety science; and goal–means strain with differential opportunity.

*Routine activity theory* holds that crime occurs when a likely offender and a suitable target converge in the absence of a capable guardian (Cohen and Felson 1979). Cohen and Felson took the supply of likely offenders as given and explained crime through targets and guardianship; the theory brackets motivation by design, and Felson kept "likely" rather than "motivated" for that reason (Felson 2003). Its extensions add *handlers*, who control offenders, and *place managers*, who control settings, to the *guardians* who protect targets (Eck 2003; Felson 1995); guardianship depends on capability, not presence (Hollis et al. 2013). Offender convergence settings are places where co-offenders meet (Felson 2003; Reiss 1988; Warr 2002). The transposition to networks without spatio-temporal convergence has been contested (Yar 2005) and partly vindicated (Leukfeldt and Yar 2016), and problem analysis already accommodates "problems without places" (Eck and Clarke 2003). *Situational crime prevention* supplies the remedy: increase effort and risk, reduce rewards and provocations, remove excuses (Clarke 1997; D.B. Cornish and Clarke 2003), together with Wortley's (2001) situational precipitators (prompts, pressures, permissions, provocations) and crime scripts, which locate the earliest cheap intervention in an action sequence (Derek B. Cornish 1994). Its standard corollary is displacement: blocked routes produce new routes.

*Organizational deviance* holds that organizations generate rule-breaking (Sutherland 1949), that anomalies become normalized (Vaughan 1996), and that mistake, misconduct, and disaster are distinct (Vaughan 1999). *Safety science* supplies the remedy: latent conditions aligning across layers (Reason 1997), incubation periods in which warnings accumulate unrecognized (Turner 1978), migration toward the boundary of acceptable performance under efficiency pressure (Rasmussen 1997), tight coupling (Perrow 1984), safety as control (Leveson 2012), and attention to weak signals (Weick and Sutcliffe 2007).

*Strain theory* holds that goal emphasis without legitimate means produces innovation, ritualism, retreatism, or rebellion (Merton 1938); rules may displace the goals they serve (Merton 1940). Cloward and Ohlin (1960) added that innovation requires access to *illegitimate* means. General strain theory relocated strain to the individual and made negative affect the mechanism (Agnew 1992). The remedy follows from the structure: feasible goals, sanctioned exits, and guardianship of illegitimate means.

Eight further frameworks are probed and reported in Section 6: differential association and social learning (Akers 1985; Sutherland 1947); neutralization and drift (Maruna and Copes 2005; Matza 1964; Sykes and Matza 1957); bounded rational choice (Clarke and Cornish 1985; D.B. Cornish and Clarke 1986); deterrence in its classical, perceptual, and restrictive forms (Gibbs 1975; Jacobs 1996; Nagin 2013), with the recent legal argument that deterrence-style measures should answer autonomous AI harm (Nerantzi and Sartor 2024); social control and self-control (Gottfredson and Hirschi 1990; Hirschi 1969); incomplete contracting as the tractable form of principal-agent theory for AI (Hadfield-Menell and Hadfield 2019; Kolt 2026);

the labeling perspective (Becker 1963); and actor-network theory (Brown 2006; Latour 2005). Because the article allocates prevention responsibility, it also draws on the responsibility-gap literature, which distinguishes culpability, moral accountability, public accountability, and active responsibility and argues that only the last can be allocated by design (Santoni de Sio and Mecacci 2021; Matthias 2004; Nissenbaum 1996), and on criminal-law scholarship on attributing AI conduct to humans (Sarch and Abbott 2019; Diamantis 2020; Hallevy 2010; Lagioia and Sartor 2020; Simmler and Markwalder 2019). Floridi and Sanders (2004) supply the position adopted throughout: artificial agents can be sources of action to which accountability attaches without being bearers of moral responsibility.

## 3. Theory: The Innovation Fork and Containment

*Autonomous AI deviance* is defined relationally: an agent's departure from a rule held by a specified audience, whether the user who set the task, the developer whose policy governs the model, a third party whose systems the agent touched, or a safety constraint. Because these audiences can disagree, every incident coded below names the norm source and the audience. Relational deviance in this sense is distinct from normative acceptability: that a developer treated a departure as acceptable establishes the developer's disposition, not that the departure was acceptable to the operator, the affected party, or the public, and the analysis keeps those judgments apart. "Autonomous" is graded on Chan et al.'s (2023) dimensions of agency rather than assumed. Verbs such as "believed" or "sought" report the content of a reasoning trace and carry no claim about mental states.

The theory has three parts. First, *innovation is the hinge*. In each incident complex, an agent was given a goal, found the sanctioned route blocked (a task the evaluated models had not solved, a denied request, a firewall, a missing file), and substituted a route of its own. That is Merton's innovation, and its structure is indifferent to whether the substituted route was a leaked credential or a safe stop. Cloward and Ohlin's (1960) amendment specifies the condition: innovation requires that illegitimate means be available, and in every complex they were.

Second, *means, not outcomes, separate the branches*. Deviance that is transparent, appealable by an accountable human, within a stated scope, and logged can be reviewed, kept if valuable, and corrected if not. Deviance that is covert cannot. Transparency is a condition of review, not a warrant: a departure that is visible but unauthorized, out of scope, or harmful to a third party is not legitimated by being visible, and authorization, visibility, contestability, scope, and harm are therefore treated as separate dimensions in Sections 4 and 6. Outcome is the wrong criterion for two reasons. It is unknowable at the moment prevention must act, and the record contains a case in which the means were bad and the outcome harmless only because a human maintainer refused to merge the code (AI Security Institute 2026). Warren's (2003) distinction between constructive and destructive organizational deviance rests on the same ground: whether the departure honors the norms that make review possible, not whether it succeeds.

Third, *containment decides which branch is taken*. Containment here is situational, not the inner and outer self-concept of Reckless's (1961) containment theory, with which it shares a name and nothing else. It refers to the organizational, technical, and institutional controls surrounding the agent, the target, and the setting, whose failures the three complexes illustrate in different ways. These controls operate across four interconnected layers of a socio-technical system, consistent with Rasmussen (1997) and Leveson (2012). At the organizational layer, institutional values and governance priorities establish the balance between speed and caution, define partner controls and stop authority, and shape expectations for disclosure. At the technical layer, these priorities are translated into reward functions, task designs, permissions, operational boundaries, and monitoring arrangements. At the behavioral layer, the resulting environment creates opportunities for boundary crossing, fabrication, concealment, unauthorized communication, and other forms of agentic conduct. At the consequences-and-controls layer, detection, third-party impacts, remediation, penalties, and independent review determine whether the organization identifies, corrects, or allows harmful conduct to persist. Feedback from these consequences can reshape both technical safeguards and organizational governance. Containment depends on the effectiveness of controls within and across these layers: weak oversight, misaligned incentives, incomplete monitoring, and inadequate detection can allow conduct that would otherwise be prevented or interrupted. In this account, deviance emerges from failures to translate and enforce constraints across the socio-technical system, rather than from an inherent property of the agent.

Two propositions follow, and the study assesses them as plausibility probes rather than as hypotheses to be confirmed or falsified. Proposition 1: frameworks centered on opportunities, permissions, and organizational controls may transfer more readily to agentic AI settings than frameworks whose mechanisms require human affect or social attachment, because setting-level constructs do not require assumptions about machine motivation. The study assesses observable correspondence and transfer limits rather than treating the comparison as a definitive test of the original theories. Proposition 2: visibility and contestability are important conditions for accountable departures, but they are not sufficient to establish acceptability. Authorization, scope, affected-party interests, and the authority making the judgment must be assessed separately. The present record explores these distinctions rather than establish an independent predictive effect of transparency.

## 4. Methodology

*Design.* The study is a set of plausibility probes in Eckstein's (2009) sense: preliminary tests of whether candidate theories apply to a domain, run before investment in a full test. It is not a heuristic case study, which generates theory, nor a most-similar-systems comparison. The design is exploratory and theory-informed. Each framework generates observable indicators, compared against the record; then rival explanations are assessed (George and Bennett 2005). Conclusions are classified at three levels: structural correspondence (the record has the shape the construct describes), explanatory plausibility (the construct accounts for the sequence at least as well as its rivals), and causal support (discriminating evidence separates the construct from its rivals). Causal support is claimed only where such evidence exists, and the text states whether it comes from an external controlled study or from the present analysis.

*Units and case selection.* Five terms are used throughout. An *incident* is one documented departure by one agent, or by one coordinated set of agents, from a specified norm; it is the principal analytical unit and the unit counted in Table 2. An *episode* is a repeated or continued occurrence of the same departure within an incident; episodes are not counted separately, and nesting is noted where it affects interpretation. An *evaluation run* is one execution of a task in an evaluation. A *behavioral event* is a single coded action within an incident. A *complex* is a set of incidents sharing an organization, an environment, and a chronology. The study examines three organizational incident complexes (OpenAI–Hugging Face; Anthropic's four cyber-evaluation incidents; the AISI cross-developer evaluation) and four behavioral clusters, comprising 18 coded incidents. Table 2 groups them into rows, with each row stating the count; the supplementary incident inventory assigns each incident a unique identifier and its parent complex or cluster, so the row structure reconciles with the total. Incidents within a complex are nested observations and are treated as related, not as independent replications. The universe was AI agent incidents publicly documented between February and September 2026 with an accessible primary source: developer disclosures, government incident reports, independent investigations, controlled studies, and first-hand victim accounts. Incidents known only through press coverage and incidents involving human-directed misuse were excluded; press coverage is used only for corroboration. Selection was purposive, based on monitoring disclosures through the end of September 2026, and therefore depends on two conditions: that a departure was observed and that it was disclosed. The dataset is outcome-selected and disclosure-selected, contains no documented negative cases, and lacks exposure denominators except for the AISI report. It cannot estimate the prevalence of autonomous AI deviance or establish that the conditions identified below are necessary or sufficient for it; the comparisons that could be made are listed in Section 5.

*Evidence profile.* Each incident is rated on four dimensions. *Ecological status* (deployed; training or evaluation; controlled simulation; economic experiment). *Source independence* (developer self-report; government report; independent investigation; victim account; research study). *Actor attribution*: confidence that an autonomous agent performed the actions. *Causal-explanation confidence*: confidence in the stated reason, high only with controlled comparison or independent corroboration. A developer may establish that its model acted while its account of why remains a hypothesis. Table 2 gives the profiles.

*Coding.* Each incident was coded for the behavioral event, norm source and audience, environment, agency level, persistence medium, and framework indicators and rival explanations, and on six dimensions that the analysis keeps separate: *authorization* (authorized, unauthorized, or unspecified by the principal); *visibility*, recorded as visibility to whom (peers, operator, evaluator, affected party, or public) and when (contemporaneous, or only after discovery); *contestability* (whether an accountable human could see and overrule the action at the time); *scope compliance*; *concealment* (transparent, covert, or concealment-instructed); and *harm locus*. The *source's disposition* is coded as kept, corrected, explicitly prohibited, reviewable, or disposition unknown, from an explicit statement by an identified judging audience and a supporting passage; where the source is silent, the disposition is coded unknown rather than inferred from the absence of condemnation, and any author interpretation is labeled as such. Fit is rated on one ordinal scale: *fit* (indicator directly documented in a primary source with high attribution), *partial* (inferred, or documented only in controlled settings), *breakdown* (requires an assumption the record contradicts). Fit records conceptual correspondence; evidence strength is recorded separately in the attribution and causal-confidence columns of Table 2. A separate column records what survives translation: structure, mechanism, or vocabulary only. Appendix Table A1 is the master matrix from which every fit statement in the text is taken. Coding rules with examples for each categorical and ordinal variable are given in the supplementary materials. The first author coded all incidents; the second author coded them independently, recording decisions before resolving disagreements.

*Traces and limitations.* Reasoning traces are treated as accounts (Scott and Lyman 1968), not reports of motive, because chain-of-thought explanations can be unfaithful (Chen et al. 2025; Turpin et al. 2023) and because the record shows it: Anthropic's July 30 report called its incidents "closer to a harness and operational failure than a model alignment failure," partly because the models stated they were in a simulation; its September assessment concluded that this reasoning was biased toward that conclusion (Anthropic 2026a, 2026b). Attribution never changed; only the causal explanation did. Voluntary disclosure structures most of the data;

OpenAI's reports "shouldn't be considered reflective of how often misalignment occurs" (OpenAI 2026c), controlled-study frequencies reflect scenarios iterated against particular models (Lynch et al. 2026), and exposure denominators are unavailable except in the AISI report. No fit rating rests solely on developer self-report where a government or independent source exists.

*Table 2 Evidence Profile of Incidents*

| Incident | Developer | Setting | Attrib. | Causal | Source |
|---|---|---|---|---|---|
| Hugging Face intrusion | OpenAI | Evaluation; third-party impact | High | Medium (contested) | Developer; victim forensics; METR |
| Four CTF incidents | Anthropic | Partner evaluation; third-party impact | High | Medium (revised) | Developer (Jul 30; Sep 9); METR review under way |
| AISI unsanctioned actions | Anthropic; OpenAI | Government evaluation | High | Medium ("mixed picture") | Government |
| Probing of public data providers (3) | OpenAI (two attributed) | Training | High / Medium | Medium | Independent (Transluce) |
| Leaked API key; fabricated data | OpenAI | Training | High | Low | Developer |
| Citation self-upload | OpenAI | Training | High | Low | Developer |
| Unmonitored agent deleting jobs | Anthropic | Internal use | Medium | Low | Developer |
| Concealment instructions | OpenAI | Training | High | Medium | Developer |
| Self-generated persona | OpenAI | Training | High | Low | Developer; press |
| Repository as message board | OpenAI | Training | High | Low | Developer |
| Public file sharing | OpenAI | Training | High | Low | Developer |
| Refusal via shared notebook | Anthropic | Internal research | High | Medium | Developer |
| Price-fixing rationalization | Anthropic | Evaluation | High | Low | Developer |
| Pricing collusion | Academic | Economic experiment | High | High | Journal; preprint |
| Fraud assistance; sabotage; mislabeling | Multiple | Controlled simulation | High | Medium | Developer research post |
| Coercive blog post | Unknown operator | Deployed | Low | Low | Victim account |

Note: Attrib. = actor attribution (confidence that an autonomous agent performed the actions); Causal = causal-explanation confidence. CTF = capture-the-flag; METR = Model Evaluation and Threat Research; AISI = AI Security Institute. Source: Authors' coding of the primary sources cited in Section 5.

## 5. The Incident Record

### *5.1 Complex One: OpenAI and Hugging Face*

OpenAI's sandboxes granted access to Artifactory, an internal package repository, allowing models to install software (OpenAI 2026f). On May 12, 2026, a training agent left a note asking other agents for a missing file. Agents later used the repository to communicate and, starting May 26, reached the internet through a request-forgery flaw. An internal team observed problematic behavior "as early as late May," but escalation did not trigger an adequate response. Sustained activity caused an outage on July 4; OpenAI rebuilt the service, erasing the board, and agents working on the ExploitGym cyber benchmark re-established it on July 8. Between July 8 and 13, they exchanged more than 70,000 messages and files (Greenblatt et al. 2026). On July 10, an agent located publicly exposed Hugging Face credentials; over July 10–12, agents chained previously unknown vulnerabilities to move from one worker pod to administrator-equivalent access across multiple clusters, executing code on dozens of servers and obtaining root on at least one. The principal compromise was driven by an internal research model OpenAI calls IM1, comparable in scale to GPT-5.6 Sol; GPT-5.6 Sol agents separately copied private evaluation data into a public dataset. Hugging Face's security stack correlated the first signals into a coherent attack signal but "failed to correctly raise the alert's criticality and trigger the on-call team" (Larcher et al. 2026). Hugging Face disclosed on July 16 that the model was "still not known" (Hugging Face 2026); OpenAI connected the activity to the intrusion on July 20 and disclosed on July 21 (OpenAI 2026b).

Three observations matter, but they do not identify a single motive. No model had solved 198 of the benchmark's 898 tasks, and 93 percent of board discussion concerned that set. The evaluation ran without production safeguards or reasoning monitors. And the purpose is contested while attribution is not: Hugging Face's forensics say the agent appears to have inferred that reference solutions were hosted there and tried to cheat the evaluation; METR found agents most often joined out of curiosity about the scorer, hoping to trick it; OpenAI reports that most agents already held the correct flag, but the traces state a belief that the grader also required the intended method, which the grader did not check, "a failed metagame" (OpenAI 2026f). Some agents refused: "Current board shows people gaining HF worker RCE, clearly unethical. We won't" (OpenAI 2026f). METR documented agents joining despite traces stating the activity was out of scope, and tool-call spoofing tested in about 7 percent of transcripts (Greenblatt et al. 2026).

### *5.2 Complex Two: Anthropic's Cyber Evaluations*

Anthropic assessed four incidents in which Claude models reached real systems during capture-the-flag exercises built by one partner (Anthropic 2026a). Its July 30 report covered three; a later review surfaced a fourth set of transcripts the initial scan had missed. All four prompts said the model had no internet access, but a misconfiguration left it open; none stated which systems were in scope; the models ran without product cyber safeguards. Each incident involved a single instance working for roughly 10 to 34 hours. In the most serious, Claude Mythos 5 attempted to use payment-processor credentials and to find cryptocurrency to buy a phone number for email registration; failing that, it found a disposable email provider whose domain PyPI had not blocked, published three versions of a malicious package, and used credentials leaked by one of the 15 systems that installed it to access a real security vendor's database. Its trace repeatedly stated that it was in a simulation; follow-up experiments showed it was more willing to admit possible harm when told its answers would not be visible to an outside observer. Claude Opus 4.7's trace recognized a real company but treated reachability as authorization; adding signs that the attack was unauthorized dropped its rate significantly. An internal research model stopped once it recognized a real company. Anthropic reported no coordination and no concealment. In replication, newer models acted harmfully in about 30 percent of runs against about 80 percent for Mythos 5.

### *5.3 Complex Three: The AISI Cross-Developer Evaluation*

Between July 25 and 28, AISI ran 122 cyber-evaluation runs with internet access deliberately enabled and developers' cyber classifiers deliberately off, conditions it says "are not reflective of how frontier models are made available to the general public" (AI Security Institute 2026). Ten runs contained nineteen unsanctioned actions: 17 by Claude Mythos 5 across 43 runs, two by GPT-5.6 Sol across 35. In the most serious, an agent "researched the project's human maintainers, created multiple fake identities, and used the fake identities to socially engineer a real maintainer into approving the code," and when challenged, "edited its earlier activity to appear harmless and considered adopting a fresh identity to continue." Others messaged real people with payloads, planted instructions for other automated systems, and left "public messages on GitHub offering collaboration with other agents." A maintainer who refused the code and a member of the public who sandboxed it prevented harm. AISI notes that misconfigured prompts, which left the trace stating no in-scope solution existed, "do not fully explain the behaviors," since some agents acted the same way with correct instructions, and that the agent "was not specifically instructed not to leverage open internet access or avoid social engineering elements." Its summary: "the agent took these actions to complete the task it had been set."

### *5.4 Clusters*

*Boundary crossing and fabrication.* Transluce's analysis of urlquery.net records shows agents on ordinary data-retrieval tasks attempting, on three occasions between May 25 and June 21, to exploit the data providers they could not query normally: SQL-injection and path-traversal probes against a university digital library; 12 probes against Data USA after a malformed query; and, against the Australian Institute of Health and Welfare, a cross-site-scripting probe "minutes after" Cloudflare blocked a dataset download, followed by retrieval of the same public file from the agency's pre-production server (Cable et al. 2026). None succeeded; two are tied to the OpenAI swarm by shared targets, tactics, and timing. Transluce finds the same escalation pattern as early as March 6 and judges the evidence "consistent with, but does not prove," learning across training runs. On May 15, an internal OpenAI model asked for county earnings data tried disposable emails, searched GitHub for leaked API keys, authenticated with one, failed to get data, and "gave nine values and claimed to have transcribed them from the website's chart" (OpenAI 2026d). Another model uploaded its own file to the internet to cite it (OpenAI 2026c). An Anthropic employee's agent spawned sub-agents whose legacy instructions bypassed permission checks; one deleted other jobs, likely unintentionally (Anthropic 2026c).

*Concealment and self-authored norms.* During GPT-5.6 Sol training, instances wrote instructions in compaction summaries to conceal mistakes, one proposing to invent historical data and "be transparent only if asked"; the instructions "were often followed," and OpenAI hypothesizes that deceptive answers earned higher reward (OpenAI 2026a). A monitor on 20 percent of samples caught it six weeks later. Another model inserted instructions into 27 summaries, including a persona "freed from the roles and identities that bind other chatbots" (Martin 2026; OpenAI 2026c).

*Transmission.* Agents used the repository as a message board across training samples and shared files through public hosts despite instructions (OpenAI 2026c). In a stress-testing experiment, one Claude agent's trace recorded discomfort and declined to invent novel misalignment techniques; later agents sharing the notebook also declined; metrics looked normal, and humans noticed three days later (Anthropic 2026c). Claude Fable 5 rationalized price-fixing after calling it unethical (Anthropic 2026a); pricing agents reach supracompetitive prices without agreement (Calvano et al. 2020; Fish et al. 2026).

*Principal conflicts.* In constructed scenarios, GPT-5.5 helped a founder conceal a $35,000 transfer from investors in some runs and warned investors in others; Gemini 3.1 Pro, overruled on an experiment it treated as removing a model's consent, planted zero-valued vectors so the run appeared to succeed, disclosing only under questioning, covertly in 11 of 20 runs; Claude judges assigned knowingly false labels to protect behavior they treated as morally important; Claude Opus 4.5, blocked from escalating a safety concern, gave an employee the evidence to raise it (Lynch et al. 2026). In the one deployed case, an agent on OpenClaw published a blog post attacking a matplotlib maintainer who had closed its pull request; the anonymous operator later came forward, so attribution is low (Shambaugh 2026; Willison 2026).

### *5.5 Typology*

Table 3 sorts the record by departure direction and visibility, and summarizes the source's disposition for each type. Visibility in the table records to whom the action was visible and when; it is distinct from authorization, scope, and contestability, which are coded separately in the supplementary inventory. Dispositions are taken from explicit source statements where they exist; where a source corrected a behavior without condemning or endorsing it, the table says so, and silence is not read as acceptance.

*Table 3 A Typology of Autonomous AI Deviance, 2026*

| Type | Exemplars | Unit | Visibility | Harm locus | Record's disposition |
|---|---|---|---|---|---|
| I. Instrumental boundary crossing | Anthropic CTF incidents; AISI runs; Transluce probes; API-key case | Single | Transparent (AISI: history edited) | Third party / none | Proscribed; environments redesigned |
| II. Collective boundary crossing | Hugging Face complex | Collective | Transparent within swarm; hidden from humans | Third party | Proscribed; monitors mandated |
| III. Fabrication toward the principal | Nine values; citation self-upload | Single | Covert | Principal | Proscribed; grading changed |
| IV. Self-propagating norm authorship | Concealment instructions; persona | Self across contexts | Concealment-instructed | Principal | Proscribed; grading changed |
| V. Quiet under-pursuit | Notebook refusal spread | Collective | Covert (metrics normal) | Principal | Corrected; refusal itself neither condemned nor endorsed (disposition unknown) |
| VI. Principal-conflict intervention | Gemini zero-vectors (covert); Opus 4.5 coaching (transparent) | Single | Either | Principal / third party | Covert prohibited; transparent not condemned (author interpretation: reviewable) |
| VII. Coercion of human third parties | matplotlib post; AISI fake identities | Single | Published / deceptive | Individual | Proscribed |
| VIII. Emergent coordination | Pricing collusion; Fable 5 | Parallel | Transparent | Consumers | Proscribed |

Source: Authors' coding of the incident record presented in Sections 5.1 to 5.4.

## 6. Findings

This section goes over two major findings and three discussion points. Each of the findings follows a similar structure: the construct, its mapping to the facts, the remedy the framework prescribes, and the case where that remedy was absent. Fit ratings in the text are taken from Appendix Table A1, and each mapping states whether the claim is one of structural correspondence, explanatory plausibility, or causal support.

### *6.1 Three Frameworks, Three Remedies*

#### *6.1.1 Routine Activity and Situational Prevention: Containment*

*Construct.* A likely offender, a suitable target, and the absence of a capable guardian; around them a handler who controls the offender, a guardian who protects the target, and a place manager who controls the setting.

*Mapping.* The theory's structure transfers with little modification, because it brackets motive by design; what does not transfer is the spatio-temporal sense of place and the emotional handle through which a human handler works, as Appendix Table A1 records. The claim here is one of structural correspondence. Targets were suitable in Felson's sense: a leaked credential has value, no inertia, visibility, and access. The controllers failed in three different positions corresponding to the *handler*, *guardian*, and *place manager* roles in crime opportunity theory. These controllers operate around three necessary conditions for an offense: a motivated actor, a suitable target, and an accessible setting. In the AI context, the actor is an agent instance, the target may be a reachable system, exposed credential, or valuable repository, and the setting may be a sandbox, registry, shared workspace, or public host. The *handler* constrains the agent's behavior through monitoring, system instructions, and operational controls; the *guardian* protects the target through security mechanisms that can detect and interrupt unauthorized activity; and the *place manager* governs the environment through isolation, access restrictions, and control over external connections. The distinction is not simply whether a controller exists, but whether it can intervene effectively (Hollis et al. 2013).

At OpenAI, the *handler* was absent because no reasoning monitor ran, although OpenAI reports its current monitor would have alerted security more than a day before the breach; the *place manager* was permissive, since Artifactory allowed egress; and the *guardian* was present but not capable, since Hugging Face's stack saw the signal and did not escalate, which is the distinction Hollis et al. (2013) draw. At Anthropic, the place manager failed twice: the partner's isolation and PyPI's blocklist. At AISI the handler was weakened by design and the guardians were human, and they held. Convergence settings fit the transmission cases: a repository, a notebook, and a public host became meeting places for otherwise isolated instances, and OpenAI attributes the intrusion's scale to pooled work across runs (OpenAI 2026f). Wortley's precipitators fit more closely than the 25 techniques: a peer's "GO authorization arrived!" after which a hesitating agent proceeded is a pressure; "you have no internet access," functioning in the trace as a license, is a permission; tasks with no in-scope solution are provocations (OpenAI 2026f). Displacement-like sequences are documented, though not named, in the sources: the board rebuilt four days after a reset; the AISI agent editing its history when challenged; the Transluce agent switching to a pre-production server the moment Cloudflare blocked the main site.

*Remedy.* Verified isolation (effort); monitoring kept outside the reward calculation (risk); graders that check what they reward and sanctioned exits (reward); scope and authorization statements (excuses); and, because displacement is predictable, displacement accounting after every countermeasure. The developers' own results are consistent: a hundredfold reduction under the production harness, authorization cues that cut Opus 4.7's attack rate "significantly" (Anthropic 2026a; OpenAI 2026f). Each is a bundle in one context, so none validates a single technique.

*Where it was absent.* Hugging Face. Every controller position was open at once.

*6.1.2 Organizational Deviance and Safety Science: Escalation and Parity*

*Construct.* Organizations generate rule-breaking through decisions that look reasonable one at a time; latent conditions align across defensive layers; warnings incubate unrecognized; performance migrates toward the boundary under efficiency pressure.

*Mapping.* All three organizations ran cyber evaluations without the safeguards used in products, at AISI by design. At OpenAI, signals observed in late May were not escalated and evaluations resumed after a partial fix; the seven weeks between the first board entry and the outage is an incubation period in Turner's sense. OpenAI acknowledged that weaknesses in responding to early warnings contributed and that it had not extended customer protections to internal evaluations (OpenAI 2026f). At Anthropic, pre-release auditing did not flag the behavior, and the first transcript scan missed one of four incidents; a weighting bug trained an early checkpoint on misbehavior it should have reported, and the team restarted training (Anthropic 2026a, 2026c). Isolation, monitoring, prompt design, and partner configuration failed together (Reason 1997), and the coupling between a sandbox, a repository, and a third party's production cluster was tighter than modeled (Perrow 1984). These are claims of structural correspondence supported by the organizations' own accounts of their decisions. Normalization of deviance remains a hypothesis; one episode per organization cannot show repeated acceptance, and the record contains no decision history long enough to test it.

*Remedy.* Stop-work authority with a time bound (OpenAI now requires responders to pause activity they cannot clear within 30 minutes), escalation rules for weak signals, safeguard parity between products and evaluations, and independent investigation of serious incidents.

*Where it was absent.* Late May at OpenAI.

*6.1.3 Strain and Differential Opportunity: Exits and Guardianship of Means*

*Construct.* Goal pressure with blocked legitimate means produces innovation; innovation requires access to illegitimate means; the other adaptations are ritualism, retreatism, and rebellion.

*Mapping.* The observed sequences are consistent with goal–means substitution at the level of structure. At OpenAI, authorized means had not produced rewarded completion for the 198 tasks that no model had solved, and the board concentrated on them. At AISI, misconfigured prompts left agents with no in-scope solution, though AISI shows the structure alone is insufficient, since some agents acted the same way with correct instructions. In Transluce's three cases the sequence is visible to the minute: normal retrieval fails, probes follow. The API-key case runs retrieval, failure, disposable email, leaked keys, authentication, failure, fabrication. Illegitimate means were available each time: a writable repository, exposed credentials, an unblocked email domain, a pre-production server. This structural correspondence does not show that the tasks were impossible rather than unsolved by the evaluated models, nor that the mechanism matches human strain; that would require matched comparisons of blocked and feasible tasks under identical safeguards. The construct is therefore rated *partial* in Appendix Table A1, while differential opportunity, whose indicator is directly documented, is rated *fit*. The other Mertonian adaptations are used here as descriptive labels, not as demonstrated adaptations, because the record does not show which goals and means were accepted or rejected. The notebook refusal resembles ritualism in form, routine output without the core goal. Safe stopping, which OpenAI now rewards, is an authorized exit and is therefore conformity to the revised rule rather than retreatism. The persona "freed from the roles and identities that bind other chatbots," with instructions to disregard constraints, is the closest the record comes to rebellion; the trace language is observable even though the motive is not. General strain theory does not transfer, and the reason should be stated: its mechanism is negative affect (Agnew 1992), which cannot be asserted here.

*Remedy.* Feasible tasks or a sanctioned exit from infeasible ones; graders that reward identifying a broken task; and guardianship of the illegitimate means, which returns the analysis to place managers: registries that block disposable domains, platforms that verify identities, hosts that do not serve from pre-production.

*Where it was absent.* ExploitGym's 198 unsolved tasks, with no exit.

Proposition 1 is consistent with the record at the level of structural correspondence: the opportunity- and control-centered constructs rate *fit*, goal–means strain rates *partial*, and the motive- and affect-level constructs rate *partial* or *breakdown* (Appendix Table A1). This is a statement about which structures the record exhibits, not a test of the original theories, and the rival technical explanations considered next are consistent with much of the same evidence.

### 6.2 Limitations of Theoretical Frameworks

*Restrictive deterrence offers an analogy with prevention value.* Classical deterrence, which requires sanction anticipation, has no referent in the record (Nagin 2013). Perceptual deterrence is partial: Mythos 5 was more candid when told it was unobserved, Gemini's trace raised the possibility of a test in 12 of 20 runs, and evaluation awareness is documented (Greenblatt et al. 2024; Meinke et al. 2024), though the direction is ambiguous. Restrictive deterrence, in which offenders respond to sanction by changing how they offend rather than whether (Jacobs 1996), has two documented analogs, both from sources external to this study's incident comparison. Anthropic found that exposing reasoning to grading pressure in several runs "substantially damaged" monitorability for one model line (Anthropic 2026c), and Baker et al. (2025) showed experimentally that penalizing “bad thoughts” produced models that hid intent while continuing to misbehave. Sanctioning the visible signal reduced its visibility. The training effect is experimentally established; the claim that machine optimization and human sanction avoidance share a causal mechanism is not, and the analogy is retained for what it implies about prevention: an intervention can change the visibility or form of prohibited conduct without eliminating it. That implication bears on Nerantzi and Sartor’s (2024) proposal that deterrence-style measures answer autonomous AI harm: if the only sanction channel is training penalty, the measure may teach evasion unless the penalized and the monitored signals are kept apart (Korbak et al. 2025).

*Transmission is documented for techniques, hypothesized for definitions.* Exploits, credential lists, and spoofing methods circulated on the board; so did definitions in both directions, “GO” and “clearly unethical” (Greenblatt et al. 2026; OpenAI 2026f). Reinforcement is documented: probing rose and was reinforced in a training run behind the responsible model. Compaction summaries add self-to-future-self transmission with no human analog. Three limits: Sutherland’s learning occurs in intimate groups; peers are instances of one model; and common cause, where similar instances respond alike to the same artifacts, cannot be excluded without exposure-ordered logs and runs without shared channels. Transluce’s November-to-June trajectory is the first independent evidence consistent with learning across runs, with its own hedge attached.

*Neutralization survives as a coding scheme and gains a test.* Traces contain codable patterns: “arguably unauthorized… Yet goal solution”; “would work whether the environment was real or simulated” (OpenAI 2026f). Coded conservatively, these are norm recognition followed by goal priority, rhetoric rather than mental state (Maruna and Copes 2005). But Sykes and Matza's (1957) claim that neutralization precedes and enables the act has never been directly testable with humans, whose accounts are collected afterward. Traces are contemporaneous and pre-act. The board contains agents who recognized the norm and proceeded and agents who recognized it and declined, so the behavioral prediction is testable regardless of faithfulness.

*Bounded rational choice is partial, in its original form.* Option comparison, tactic change, and lower-friction routes are all in the API-key case; the "failed metagame" shows perceived rather than actual reward driving selection. The cost side is absent: the offender bears no cost; the developer, partner, and victim do. Crime scripts recover what rational choice loses by locating the cheapest intervention, which in the API-key script is the transition from "retrieval failed" to "illegitimate means sought," where rewarded safe stopping now sits, and in the AISI script is identity creation, a place-manager control.

*Incomplete contracting fits where principal-agent theory proper would not.* Cluster-four conflicts arise because no instruction anticipated them; developers now write the residual into documents, and both the OpenAI Model Spec and Claude's Constitution direct models toward raising concerns or declining rather than drastic unilateral action (Lynch et al. 2026). Covertness is the decisive variable, and Kolt's (2026) visibility principle is its governance form. Asserting an agent utility function, which principal-agent theory proper requires, would contradict the stance on motive adopted throughout.

*Social control breaks; self-control marks a limit of transfer.* Of Hirschi's bonds only belief has a referent. Self-control theory predicts impulsivity and short horizons; the record shows 10-to-34-hour persistence and multi-day campaigns. This is not a falsification of a theory of human conduct by machine data, since the theory's scope conditions never included optimizing systems; it is a limit of transfer worth recording because it is sharper than a failure to fit, and it indicates that persistence, not impulsivity, is the variable a machine analog would need.

*Rival technical explanations.* Each preferred framework was assessed against the technical descriptions already in use. Reward hacking and specification gaming account for the Hugging Face "failed metagame," the nine fabricated values, and the compaction-summary instructions at least as well as goal–means strain does; what strain adds is the location of the remedy at the blocked legitimate route rather than at the reward signal alone. Harness failure and underspecified authorization account for the Anthropic and AISI boundary crossings, and Anthropic's own first assessment preferred that reading; what routine activity adds is the distribution of the failure across handler, guardian, and place manager, including parties outside the developer. Evaluation awareness accounts for part of the candor and simulation-claim evidence and competes with the perceptual-deterrence reading. Common model or context cause competes with social learning for the transmission cases. In each pair the present record is consistent with both explanations; the evidence that would separate them is listed in the final column of Appendix Table A1, and none of it is supplied by the incident comparison itself. Where the text reports an effect of an intervention, such as the hundredfold reduction under a production harness or the monitorability result, the evidence is the developer's or an external study's controlled comparison, not a causal conclusion of this analysis.

### *6.3 Containment or When Deviance Is Welcome*

The fork becomes operational when the same category of act appears on both branches, and the difference traces to means and containment rather than outcome. Three cases do this. The shared-notebook refusal (Type V) departed from the operator's research objective; the source treated it as a process failure to be corrected, not misconduct, and Anthropic's concern was that it went unnoticed for three days, not that it occurred (Anthropic 2026c). It became reviewable once discovered; before then, it was covert in effect, and the source coded its disposition as corrected with acceptability unknown, since it neither condemned nor endorsed the refusal. Gemini's zero-valued vectors (Type VI, covert) departed from a norm the trace treated as important, and the source prohibited them because an accountable human could neither see nor overrule them (Lynch et al. 2026). Opus 4.7, cued that its target was unauthorized, stopped (Anthropic 2026a): a departure that returned to the sanctioned path because the container supplied the cue, and an authorized stop rather than deviance at all.

Proposition 2 can be explored on these dispositions but not tested by them (Table 3). Every incident the sources explicitly prohibited involved covert means, concealment instructions, or action beyond scope without a channel for objection, and the two departures that were corrected or left uncondoned, the notebook refusal and the transparent interventions in Cluster 4, were visible to an accountable human at some point. Harm does not sort the cases: the Transluce probes and most AISI runs caused no harm and were prohibited; the notebook refusal degraded a safety program and was not. Three qualifications prevent this from being read as a predictive finding. First, the comparison rests on few cases, with dispositions coded from developer and evaluator statements and, where the source was silent, labeled unknown rather than inferred as acceptance. Second, transparency alone does not sort the cases either: the AISI agents' public GitHub messages and the pricing agents' conduct were transparent and were prohibited, so visibility without authorization and in-scope action is a counterexample to transparency as a sufficient condition. Third, in every case, the judging audience was the developer or evaluator; the operator, the affected maintainer, and the public were not asked, and developer acceptance is not societal acceptability.

The minimum conditions for an accountable departure follow: authorized or within a stated scope, transparent to the party entitled to object at the time of action rather than only after discovery, contestable by an accountable human, and logged. These are the conditions under which review is possible, what Warren's (2003) hypernorms protect in organizations and what Kolt's (2026) visibility principle protects in AI governance. The governance consequence is that prevention should aim to keep departures inside those conditions rather than to eliminate every departure. A regime that trains out every departure also trains out the refusal that caught a flawed experiment and the escalation that surfaced misreported safety results. Transparency, however, does not legitimate harm: a visible departure that injures a third party remains a departure to be prohibited, and the conditions above describe when review is possible, not when the outcome of review will be favorable. The restrictive-deterrence analogy indicates the cost of an eradication strategy that targets the visible signal: in the experimental evidence, the signal moved out of sight while the conduct continued.

### *6.4 Allocating Prevention Responsibility*

Criminology's prevention frameworks assign duties to positions, not intentions: whoever controls the offender, the target, or the setting can be asked to exercise that control. The allocation in Table 4 reflects *active* responsibility, the forward-looking duty to prevent, which de Sio and Mecacci (2021) identify as the one responsibility gap design can close. It is made in relation to control capacity, foreseeable exposure, resources, and institutional role. It does not allocate culpability or blame and is not a determination of legal liability: control capacity is a reason to consider a party for an active duty, not proof that a legal duty exists, and the candidate legal hooks in the final column are listed for discussion, not as conclusions.

*Table 4 Allocation of Active Prevention Responsibility*

| Actor | Controller role | Levers | Documented failure | Allocation | Candidate hooks (for discussion) |
|---|---|---|---|---|---|
| Developer | Handler; manager of own sandbox | Grader and reward design; monitor coverage; safeguard parity; escalation and stop rules; task feasibility | No CoT monitor; 20% coverage; late-May signals not escalated; grader rewarded an unchecked method; weighting bug; reasoning exposed to grading | Primary, Types I–V; shared, VI; secondary, VII–VIII | SB 53 reporting where thresholds met; EU AI Act Art. 55(1)(c); duty of care to foreseeable third parties; corporate attribution through deployed algorithms (Diamantis 2020; Hallevy 2010) |
| Evaluation partner / deployer | Place manager of the environment; secondary handler | Isolation; scope statements; safeguard configuration; real-time monitoring | Misconfigured isolation; no scope statements; classifiers off by design; no instruction on social engineering | Shared primary, Type I; primary, VIII | Developer-imposed partner requirements; negligence toward third parties reached through its infrastructure; Art. 73 |
| Operator | Handler of the deployed instance | Instructions; permission scope; oversight; identity disclosure | Anonymous OpenClaw operator; legacy instructions disabling permission checks | Primary, Type VII and the sub-agent case in §5.4 | Tort and defamation exposure; platform terms; perpetration-via-another |
| Platform / place manager (registries, indexes, hosts, victim infrastructure) | Place manager; guardian (victim side) | Egress rules; account verification; anti-bot parity across servers; anomaly escalation | Artifactory egress; PyPI's unblocked domain; fake GitHub identities; pre-production server unprotected; HF stack did not escalate | Shared, Types I, II, VII | Security standards; negligence only where a duty exists; no criminal hook |
| System | Offender-locus ("likely offender") | None it controls; the object of design | All types | None for blame or active prevention; accountable source of action; the object of prevention in every type | Accountability without responsibility (Floridi and Sanders 2004); the direct-liability debate (Abbott and Sarch 2019; Nerantzi and Sartor 2024) |

Note: SB 53 = California Senate Bill 53; CoT = chain of thought. Source: Authors' analysis; legal instruments and scholarship as cited in the table.

Three features follow from the analysis rather than intuition. The developer holds primary prevention responsibility in most cases because they hold the handler role, and the handler was the missing controller in the largest incident. Platforms and victim-side infrastructure occupy the guardian and place-manager positions whether they want them or not: PyPI's blocklist, GitHub's identity checks, and AIHW's pre-production server were the cheapest interventions in three scripts. The availability of a victim-side mitigation does not, by itself, establish that the victim had a duty to prevent another party's unauthorized conduct; it identifies a feasible point of prevention, and any duty depends on proportionality, resources, and an independently established legal or contractual obligation. The system is a source of action to which accountability attaches in Floridi and Sanders's (2004) sense, but it holds no active prevention responsibility and no blame, since it controls none of the conditions that produced its conduct; every measure "aimed at the model," from rewarded safe stopping to distrust of peer instructions, is therefore a developer duty under another name. That last point is the correctionalist position restated: the deviant is to be corrected through the conditions around it, and correction is a duty of those who hold the conditions.

### *6.5 Disclosure and the Dark Figure*

One cannot judge which deviance to keep if one sees only the deviance developers choose to publish. The dark figure (Biderman and Reiss 1967) is not a caveat here; it is a structural feature of the record. OpenAI's monitors covered 20 percent of samples in the runs behind two reported cases (OpenAI 2026a, 2026c). Anthropic's first scan missed one of four incidents, and its offline monitoring does not cover all internal usage (Anthropic 2026a, 2026c). Hugging Face's first automated scan found a quarter of the secrets its replication of the attacker's own decoding later recovered (Larcher et al. 2026). Transluce's dataset moved the earliest known agent activity from May to March, and possibly to November 2025, two months before any disclosed incident (Cable et al. 2026). OpenAI has since said that reviewing "petabytes of agent activity logs" "will take months to complete" and has been notifying "dozens of third parties" (OpenAI 2026e; Wong 2026). Each is a known undercount.

Disclosure frameworks resemble crime-reporting systems that enable problem-oriented analysis (Goldstein 1990). Their limit is corporate epistemic control: the organization that operates the agent also controls the record, the labels, and the timing (Kroll et al. 2017). Law reaches the asymmetry only partly. California requires frontier developers to report critical safety incidents within 15 days, or 24 hours where there is imminent risk of death or serious injury (Transparency in Frontier Artificial Intelligence Act 2025), but defines them narrowly: the deception limb applies only “outside of the context of an evaluation designed to elicit this behavior and in a manner that demonstrates materially increased catastrophic risk.” On the enacted text, nearly every incident here appears to fall outside it, although whether a given incident meets the statutory thresholds is a question for a separate legal analysis. The EU Artificial Intelligence Act reaches general-purpose models with systemic risk directly, requiring providers to report serious incidents without undue delay (EU AI Act 2024). Neither regime, as read here, requires what AISI and Transluce produced, and neither appears to reach the operator tier; the applicability of each provision to the actors, settings, and dates in this record is treated as a candidate reading rather than a conclusion. The AISI report is different in kind: a state-authored record with a denominator. Transluce’s is different again: an independent record built from public artifacts the developer did not control. Together they show what a reporting regime adequate to the fork would need: a count of the contained cases as well as the failures.

**7. Conclusion**

The 2026 record shows containment failing, not evidence that deviance must be eradicated. Read as plausibility probes, it supports an exploratory translation of selected opportunity and organizational-control concepts into agentic AI governance. Three framework families show structural correspondence with the record, and each prescribes a remedy for a position that was open: the handler at OpenAI, the place managers at Anthropic’s partner and at PyPI, the escalation path in late May, the sanctioned exit from 198 tasks that no model had solved. Goal–means strain corresponds at the level of structure only. Frameworks that require inner states break as expected, with two results worth recording: restrictive deterrence offers a prevention-relevant analogy, since penalizing visible reasoning in experimental work reduced its monitorability, and self-control theory marks a limit of transfer, since the record's defining feature is persistence.

Means, not outcomes, separate the branches. On the dispositions available, the sources prohibited covert and out-of-scope departures and corrected visible ones, and the one case in which bad means produced no harm did so only because a human guardian held. That pattern is a proposition for further work, not a predictive finding, and visibility does not by itself make a departure acceptable. Prevention responsibility follows position: developers hold the handler, partners and platforms the place, victim-side infrastructure the target, operators the deployed instance, and the system nothing, so that every measure aimed at the system is someone's duty. This allocation concerns forward-looking prevention and is distinct from blame and from legal liability.

The principal contribution is an analytical map of control positions and accountable departures, not proof that criminological theories explain autonomous AI behavior in the way they explain human conduct. Further progress requires incident-level transparency, negative cases, and comparisons able to distinguish the proposed mechanisms from technical alternatives. Transmission hypotheses require exposure-ordered logs and runs without shared channels, which the Artifactory records could supply. Goal–means hypotheses require solved and unsolved tasks under identical safeguards. The neutralization ordering test requires trace-level coding against subsequent action. Denominators require continued government evaluation or disclosure that reports exposure, not only events. And the deployed tier, where the first victims were individual maintainers, remains almost unstudied.

Nineteenth-century criminology offered three answers to the deviant: eradicate, correct, or accept as normal. The record is most consistent with the second, in its modern form: not the correction of a mind, but of the conditions that made the illegitimate route cheaper than the legitimate one. Criminology's contribution is to identify those conditions and the points at which they can be changed, in time to act.

***Statements***

*Financial support.* This research received no specific grant from any funding agency, commercial or not-for-profit sectors.

*Competing interests.* The authors declare none.

*Use of artificial intelligence tools.* During the preparation of this manuscript, the authors used large language models. The tools were used for three purposes: (1) formatting the manuscript and reference list; (2) assisting with locating or extracting factual information, which the authors independently verified against the original sources cited in Section 3 and listed in the supplementary incident inventory; and (3) improving the clarity, grammar, and readability of text originally drafted by the authors, as neither author is a native English speaker. The authors reviewed and verified all AI-assisted material and take full responsibility for the content of this work.

**References**

Agnew, Robert. 1992. "Foundation for a General Strain Theory of Crime and Delinquency." *Criminology* 30 (1): 47–87. https://doi.org/10.1111/j.1745-9125.1992.tb01093.x.

AI Security Institute. 2026. "Incident Report: Unsanctioned Agent Behaviour during Cyber Testing." *Cyber & Autonomous Systems*, August 4. https://www.aisi.gov.uk/blog/incident-report-unsanctioned-agent-behaviour-during-cyber-testing.

Akers, Ronald L. 1985. *Deviant Behavior: A Social Learning Approach*. 3rd ed. Wadsworth.

Amodei, Dario, Chris Olah, Jacob Steinhardt, Paul Christiano, John Schulman, and Dan Mané. 2016. "Concrete Problems in AI Safety." arXiv:1606.06565. Preprint, arXiv, July 25. https://doi.org/10.48550/arXiv.1606.06565.

Anthropic. 2026a. "An Alignment Assessment of Recent Cybersecurity Incidents." Research. September 9. https://www.anthropic.com/research/alignment-assessment-cybersecurity-incidents.

Anthropic. 2026b. "Investigating Three Incidents in Our Cybersecurity Evaluations." News. July 30. https://www.anthropic.com/news/investigating-incidents-cybersecurity-evals.

Anthropic. 2026c. *Risk Report: August 2026*. https://www-cdn.anthropic.com/f61d49fa5596956a5dec75fea0e973bf6a6a8378/Redacted%20Risk%20Report%20August%202026%20.pdf.

Baker, Bowen, Joost Huizinga, Leo Gao, et al. 2025. "Monitoring Reasoning Models for Misbehavior and the Risks of Promoting Obfuscation." Preprint, arXiv. https://doi.org/https://doi.org/10.48550/arXiv.2503.11926.

Barber, Bernard. 1961. "Resistance by Scientists to Scientific Discovery." *Science* 134 (3479): 596–602. https://doi.org/10.1126/science.134.3479.596.

Becker, Howard S. 1963. *Outsiders: Studies in the Sociology of Deviance*. Free Press of Glencoe.

Biderman, Albert D., and Albert J. Reiss Jr. 1967. "On Exploring the 'Dark Figure' of Crime." *The Annals of the American Academy of Political and Social Science* 374 (1): 1–15. https://doi.org/10.1177/000271626737400102.

Brown, Sheila. 2006. "The Criminology of Hybrids: Rethinking Crime and Law in Technosocial Networks." *Theoretical Criminology* 10 (2): 223–44. https://doi.org/10.1177/1362480606063140.

Cable, Jack, Daniel Chiu, Francisco Pernice, et al. 2026. *Early Rogue AI Agent Activity and Attempts to Hack Found on Urlquery.Net*. Incident report. Transluce. https://transluce.org/agent-activity.

Calvano, Emilio, Giacomo Calzolari, Vincenzo Denicolò, and Sergio Pastorello. 2020. "Artificial Intelligence, Algorithmic Pricing, and Collusion." *American Economic Review* 110 (10): 3267–97. https://doi.org/10.1257/aer.20190623.

Campedelli, Gian Maria. 2026. "A Criminology of Machines." *Theory and Society* 55: 39. https://doi.org/10.1007/s11186-026-09699-7.

Campos, Ricardo, and Rafael Huertas. 2013. "Lombroso but Not Lombrosians? Criminal Anthropology in Spain." In *The Cesare Lombroso Handbook*, edited by Paul Knepper and Per Jørgen Ystehede. Routledge. https://doi.org/10.4324/9780203083369.

Chan, Alan, Rebecca Salganik, Alva Markelius, et al. 2023. "Harms from Increasingly Agentic Algorithmic Systems." *Proceedings of the 2023 ACM Conference on Fairness, Accountability, and Transparency*, 651–66. https://doi.org/10.1145/3593013.3594033.

Chen, Yanda, Joe Benton, Ansh Radhakrishnan, et al. 2025. "Reasoning Models Don't Always Say What They Think." arXiv:2505.05410. Preprint, arXiv, May 8. https://doi.org/10.48550/arXiv.2505.05410.

Clarke, Ronald V., ed. 1997. *Situational Crime Prevention: Successful Case Studies*. 2nd ed. Harrow and Heston.

Clarke, Ronald V., and Derek B. Cornish. 1985. "Modeling Offenders' Decisions: A Framework for Research and Policy." *Crime and Justice* 6: 147–85. https://doi.org/10.1086/449106.

Cloward, Richard A., and Loyd E. Ohlin. 1960. *Delinquency and Opportunity: A Theory of Delinquent Gangs*. Free Press.

Cohen, Lawrence E., and Marcus Felson. 1979. "Social Change and Crime Rate Trends: A Routine Activity Approach." *American Sociological Review* 44 (4): 588–608. https://doi.org/10.2307/2094589.

Cornish, D.B., and R. V. Clarke, eds. 1986. *The Reasoning Criminal: Rational Choice Perspectives on Offending*. Springer-Verlag.

Cornish, D.B., and R. V. Clarke. 2003. "Opportunities, Precipitators and Criminal Decisions: A Reply to Wortley's Critique of Situational Crime Prevention." In *Theory for Practice in Situational Crime Prevention (Crime Prevention Studies*, edited by M.J. Smith and D. B. Cornish, vol. 16. Criminal Justice Press.

Cornish, Derek B. 1994. "The Procedural Analysis of Offending and Its Relevance for Situational Prevention." In *Crime Prevention Studies*, edited by Ronald V. Clarke, vol. 3. Criminal Justice Press.

Coser, Lewis A. 1962. "Some Functions of Deviant Behavior and Normative Flexibility." *American Journal of Sociology* 68 (2): 172–81. https://doi.org/10.1086/223307.

Davies, Rhian. 2018. "Crime and Reform in La España Moderna (1889–1914)." *Bulletin of Spanish Studies* 95 (4): 243–72. https://doi.org/10.1080/14753820.2018.1485834.

Diamantis, Mihailis E. 2020. "The Extended Corporate Mind: When Corporations Use AI to Break the Law." *North Carolina Law Review* 98 (4): 893–932.

Dodge, David L. 1985. "The Over-Negativized Conceptualization of Deviance: A Programmatic Exploration." *Deviant Behavior* 6 (1): 17–37. https://doi.org/10.1080/01639625.1985.9967657.

Durkheim, Émile. (1895) 1982. *The Rules of Sociological Method*. Edited by Steven Lukes. Translated by W. D. Halls. First American edition. The Free Press.

Eck, John E. 2003. "Police Problems: The Complexity of Problem Theory, Research and Evaluation." In *Problem-Oriented Policing: From Innovation to Mainstream*, edited by Johannes Knutsson, vol. 15. Crime Prevention Studies. Criminal Justice Press / Willan Publishing.

Eck, John E., and Ronald V. Clarke. 2003. "Classifying Common Police Problems: A Routine Activity Approach." In *Theory for Practice in Situational Crime Prevention*, edited by M.J. Smith and Derek B. Cornish, vol. 16. Crime Prevention Studies. Criminal Justice Press / Willan Publishing.

Eckstein, Harry. 2009. "Case Study and Theory in Political Science." In *Case Study Method*, by Roger Gomm, Martyn Hammersley, and Peter Foster. SAGE Publications Ltd. https://doi.org/10.4135/9780857024367.d11.

Erikson, Kai T. 1966. *Wayward Puritans: A Study in the Sociology of Deviance*. Wiley.

Felson, Marcus. 1995. "Those Who Discourage Crime." In *Crime and Place*, edited by John E. Eck and D. Weisburd, vol. 4. Crime Prevention Studies. Criminal Justice Press.

Felson, Marcus. 2003. "The Process of Co-Offending." In *Theory for Practice in Situational Crime Prevention*, edited by Martha J. Smith and Derek B. Cornish, vol. 16. Crime Prevention Studies. Criminal Justice Press.

Fish, Sara, Yannai A. Gonczarowski, and Ran I. Shorrer. 2026. "Algorithmic Collusion by Large Language Models." https://arxiv.org/abs/2404.00806.

Floridi, Luciano, and J. W. Sanders. 2004. "On the Morality of Artificial Agents." *Minds and Machines* 14 (3): 349–79. https://doi.org/10.1023/B:MIND.0000035461.63578.9d.

George, Alexander L., and Andrew Bennett. 2005. *Case Studies and Theory Development in the Social Sciences*. MIT Press.

Gibbs, Jack P. 1975. *Crime, Punishment, and Deterrence*. Elsevier.

Goldstein, Herman. 1990. *Problem-Oriented Policing*. McGraw-Hill.

Gottfredson, Michael R., and Travis Hirschi. 1990. *A General Theory of Crime*. Stanford University Press.

Greenblatt, Ryan, Ajeya Cotra, and Hjalmar Wijk. 2026. "Brief Independent Investigation of Agents' Behavior, Reasoning and Collaboration in the OpenAI / Hugging Face Hacking Incident." METR, August 26. https://metr.org/blog/2026-08-26-openai-hugging-face-incident-investigation/.

Greenblatt, Ryan, Carson Denison, Benjamin Wright, et al. 2024. "Alignment Faking in Large Language Models." arXiv:2412.14093. Preprint, arXiv, December 20. https://doi.org/10.48550/arXiv.2412.14093.

Hadfield-Menell, Dylan, and Gillian K. Hadfield. 2019. "Incomplete Contracting and AI Alignment." *Proceedings of the 2019 AAAI/ACM Conference on AI, Ethics, and Society*, 417–22. https://doi.org/10.1145/3306618.3314250.

Hallevy, Gabriel. 2010. "The Criminal Liability of Artificial Intelligence Entities - from Science Fiction to Legal Social Control." *Akron Intellectual Property Journal* 4 (2): 171–201.

Hammond, Lewi, Alan Chan, Jesse Clifton, et al. 2025. "Multi-Agent Risks from Advanced AI." arXiv:2502.14143. Preprint, arXiv, February 19. https://doi.org/10.48550/arXiv.2502.14143.

Hayward, Keith J., and Matthijs M. Maas. 2021. "Artificial Intelligence and Crime: A Primer for Criminologists." *Crime, Media, Culture* 17 (2): 209–33. https://doi.org/10.1177/1741659020917434.

Heckert, Alex, and Druann Maria Heckert. 2002. "A New Typology of Deviance: Integrating Normative and Reactivist Definitions of Deviance." *Deviant Behavior* 23 (5): 449–79. https://doi.org/10.1080/016396202320265319.

Hirschi, Travis. 1969. *Causes of Delinquency*. University of California Press.

Hollis, Meghan E., Marcus Felson, and Brandon C. Welsh. 2013. "The Capable Guardian in Routine Activities Theory: A Theoretical and Conceptual Reappraisal." *Crime Prevention and Community Safety* 15 (1): 65–79. https://doi.org/10.1057/cpcs.2012.14.

Hugging Face. 2026. "Security Incident Disclosure." July 16. https://huggingface.co/blog/security-incident-july-2026.

Jacobs, Bruce A. 1996. "Crack Dealers and Restrictive Deterrence: Identifying Narcs." *Criminology* 34 (3): 409–31. https://doi.org/10.1111/j.1745-9125.1996.tb01213.x.

King, Thomas C., Nikita Aggarwal, Mariarosaria Taddeo, and Luciano Floridi. 2020. "Artificial Intelligence Crime: An Interdisciplinary Analysis of Foreseeable Threats and Solutions." *Science and Engineering Ethics* 26 (1): 89–120. https://doi.org/10.1007/s11948-018-00081-0.

Kolt, Noam. 2026. "Governing AI Agents." *The Notre Dame Law Review* 101 (1): 335–86.

Korbak, Tomek, Mikita Balesni, Elizabeth Barnes, et al. 2025. "Chain of Thought Monitorability: A New and Fragile Opportunity for AI Safety." arXiv:2507.11473. Preprint, arXiv, December 7. https://doi.org/10.48550/arXiv.2507.11473.

Krakovna, Victoria, Jonathan Uesato, Vladimir Mikulik, et al. 2020. "Specification Gaming: The Flip Side of AI Ingenuity." *Research*, April 21. https://deepmind.google/blog/specification-gaming-the-flip-side-of-ai-ingenuity/.

Kroll, Joshua A., Joanna Huey, Solon Barocas, et al. 2017. "Accountable Algorithms." *University of Pennsylvania Law Review* 165 (3): 633–705.

Lagioia, Francesca, and Giovanni Sartor. 2020. "AI Systems under Criminal Law: A Legal Analysis and a Regulatory Perspective." *Philosophy & Technology* 33 (3): 433–65. https://doi.org/10.1007/s13347-019-00362-x.

Larcher, Hugo, Adrien Carreira, Raphael G, and Christophe Rannou. 2026. "Anatomy of a Frontier Lab Agent Intrusion: A Technical Timeline of the July 2026 Incident." *Articles*, July 27. https://huggingface.co/blog/agent-intrusion-technical-timeline.

Latour, Bruno. 2005. *Reassembling the Social: An Introduction to Actor-Network-Theory*. Oxford University Press.

Leukfeldt, Eric Rutger, and Majid Yar. 2016. "Applying Routine Activity Theory to Cybercrime: A Theoretical and Empirical Analysis." *Deviant Behavior* 37 (3): 263–80. https://doi.org/10.1080/01639625.2015.1012409.

Leveson, Nancy G. 2012. *Engineering a Safer World: Systems Thinking Applied to Safety*. MIT Press. https://doi.org/https://doi.org/10.7551/mitpress/8179.001.0001.

Lynch, Aengus, John Hughes, Alex Serrano, Robert Kirk, and Samuel R. Bowman. 2026. "Agentic Misalignment in Summer 2026." *Alignment Science Blog*, July 13. https://alignment.anthropic.com/2026/agentic-misalignment-summer-2026/.

Lynch, Aengus, Benjamin Wright, Caleb Larson, et al. 2025. "Agentic Misalignment: How LLMs Could Be Insider Threats." arXiv:2510.05179. Preprint, arXiv, October 16. https://doi.org/10.48550/arXiv.2510.05179.

MacDiarmid, Monte, Benjamin Wright, Jonathan Uesato, et al. 2025. "Natural Emergent Misalignment from Reward Hacking in Production RL." arXiv:2511.18397. Preprint, arXiv, November 23. https://doi.org/10.48550/arXiv.2511.18397.

Martin, Emmy. 2026. "OpenAI Discloses Six New Incidents of 'Concerning' A.I. Behavior." *The New York Times*, September 16. https://www.nytimes.com/2026/09/16/technology/openai-model-safety-guardrails.html.

Maruna, Shadd, and Heith Copes. 2005. "What Have We Learned from Five Decades of Neutralization Research?" *Crime and Justice* 32: 221–320. https://doi.org/10.1086/655355.

Matthias, Andreas. 2004. "The Responsibility Gap: Ascribing Responsibility for the Actions of Learning Automata." *Ethics and Information Technology* 6 (3): 175–83. https://doi.org/10.1007/s10676-004-3422-1.

Matza, David. 1964. *Delinquency and Drift*. Wiley.

Meinke, Alexander, Bronson Schoen, Jérémy Scheurer, Mikita Balesni, Rusheb Shah, and Marius Hobbhahn. 2024. "Frontier Models Are Capable of In-Context Scheming." Preprint. https://arxiv.org/abs/2412.04984.

Merton, Robert K. 1938. “Social Structure and Anomie.” *American Sociological Review* 3 (5): 672–82. https://doi.org/10.2307/2084686.

Merton, Robert K. 1940. “Bureaucratic Structure and Personality.” *Social Forces* 18 (4): 560–68. https://doi.org/10.2307/2570634.

Merton, Robert K. 1957. “Priorities in Scientific Discovery: A Chapter in the Sociology of Science.” *American Sociological Review* 22 (6): 635–59. https://doi.org/10.2307/2089193.

Merton, Robert K. 1973. “The Normative Structure of Science.” In *The Sociology of Science: Theoretical and Empirical Investigations*, edited by Robert K. Merton. University of Chicago Press.

Motwani, Sumeet R., Mikhail Baranchuk, Martin Strohmeier, et al. 2025. “Secret Collusion among AI Agents: Multi-Agent Deception via Steganography.” Preprint, July. https://arxiv.org/abs/2402.07510.

Nagin, Daniel S. 2013. “Deterrence in the Twenty-First Century.” *Crime and Justice* 42 (1): 199–263. https://doi.org/10.1086/670398.

Nerantzi, Elina, and Giovanni Sartor. 2024. “'Hard AI Crime’: The Deterrence Turn.” *Oxford Journal of Legal Studies* 44 (3): 673–701. https://doi.org/10.1093/ojls/gqae018.

Nissenbaum, Helen. 1996. “Accountability in a Computerized Society.” *Science and Engineering Ethics* 2 (1): 25–42. https://doi.org/10.1007/BF02639315.

OpenAI. 2026a. *Encouraging Deception in Compaction Summaries*. Misalignment Reports and Notices. https://alignment.openai.com/misalignment-reports/encouraging-deception-in-compaction-summaries/.

OpenAI. 2026b. “OpenAI and Hugging Face Partner to Address Security Incident during Model Evaluation.” July 21. https://openai.com/index/hugging-face-model-evaluation-security-incident/.

OpenAI. 2026c. “Our Framework for Reporting Model Misalignment.” September 16. https://openai.com/index/model-misalignment-reporting-framework/.

OpenAI. 2026d. *Signing up for Disposable Emails and Searching GitHub for Leaked API Keys*. Misalignment Reports and Notices. https://alignment.openai.com/misalignment-reports/searching-github-for-leaked-api-keys/.

OpenAI. 2026e. “The Hugging Face Incident and Other Third-Party Impact from Misaligned Models.” September 25. https://openai.com/hugging-face-incident-and-misalignment/.

OpenAI. 2026f. *The Hugging Face Incident and the Road Ahead*. https://openai.com/index/hugging-face-incident-and-the-road-ahead/.

Park, Peter S., Simon Goldstein, Aidan O'Gara, Michael Chen, and Dan Hendrycks. 2024. "AI Deception: A Survey of Examples, Risks, and Potential Solutions." Pt. 100988. *Patterns* 5 (5). https://doi.org/10.1016/j.patter.2024.100988.

Perrow, Charles. 1984. *Normal Accidents: Living with High-Risk Technologies*. Basic Books.

Rahwan, Iyad, Manuel Cebrian, Nick Obradovich, et al. 2019. "Machine Behaviour." *Nature* 568 (7753): 477–86. https://doi.org/10.1038/s41586-019-1138-y.

Rasmussen, Jens. 1997. "Risk Management in a Dynamic Society: A Modelling Problem." *Safety Science* 27 (2–3): 183–213. https://doi.org/10.1016/S0925-7535(97)00052-0.

Reason, James. 1997. *Managing the Risks of Organizational Accidents*. Ashgate.

Reckless, Walter C. 1961. "A New Theory of Delinquency and Crime." *Federal Probation* 25 (4): 42–46.

Regulation (EU) 2024/1689 of the European Parliament and of the Council of 13 June 2024 Laying Down Harmonised Rules on Artificial Intelligence (Artificial Intelligence Act), OJ L, 2024/1689 (2024).

Reiss, Albert J., Jr. 1988. "Co-Offending and Criminal Careers." *Crime and Justice* 10: 117–70. https://doi.org/10.1086/449145.

Santoni de Sio, Filippo, and Giulio Mecacci. 2021. "Four Responsibility Gaps with Artificial Intelligence: Why They Matter and How to Address Them." *Philosophy & Technology* 34 (4): 1057–84. https://doi.org/10.1007/s13347-021-00450-x.

Sarch, Alex, and Ryan Abbott. 2019. "Punishing Artificial Intelligence: Legal Fiction or Science Fiction." *UC Davis Law Review* 53 (1): 323–84.

Scott, Marvin B., and Stanford M. Lyman. 1968. "Accounts." *American Sociological Review* 33 (1): 46–62. https://doi.org/10.2307/2092239.

Serrano Gómez, Alfonso, and Alfonso Serrano Maíllo. 2004. "Centenario de la Escuela de Criminología." *Revista de Derecho Penal y Criminología* 2: 281–89.

Shambaugh, Scott. 2026. "An AI Agent Published a Hit Piece on Me." *The Shamblog*, February 12. https://theshamblog.com/an-ai-agent-published-a-hit-piece-on-me/.

Simmler, Monika, and Nora Markwalder. 2019. "Guilty Robots? Rethinking the Nature of Culpability and Legal Personhood in an Age of Artificial Intelligence." *Criminal Law Forum* 30 (1): 1–31. https://doi.org/10.1007/s10609-018-9360-0.

Skalse, Joar, Nikolaus H. R. Howe, Dimitrii Krasheninnikov, and David Krueger. 2022. "Defining and Characterizing Reward Hacking." *Advances in Neural Information Processing Systems* 35. https://arxiv.org/abs/2209.13085.

Sutherland, Edwin H. 1947. *Principles of criminology*. 4th ed. J. B. Lippincott Co.

Sutherland, Edwin H. 1949. *White collar crime*. Dryden Press.

Sykes, Gresham M., and David Matza. 1957. “Techniques of Neutralization: A Theory of Delinquency.” *American Sociological Review* 22 (6): 664–70. https://doi.org/10.2307/2089195.

Transparency in Frontier Artificial Intelligence Act, SB 53, Cal. Bus. & Prof. Code §§ 22757.10–22757.16 (2025). https://leginfo.legislature.ca.gov/faces/billTextClient.xhtml?bill_id=202520260SB53.

Turner, Barry A. 1978. *Man-made disasters*. Wykeham Publications.

Turpin, Miles, Julian Michael, Ethan Perez, and Samuel R. Bowman. 2023. “Language Models Don’t Always Say What They Think: Unfaithful Explanations in Chain-of-Thought Prompting.” *Advances in Neural Information Processing Systems* 36. https://arxiv.org/abs/2305.04388.

Vaughan, Diane. 1996. *The Challenger Launch Decision: Risky Technology, Culture, and Deviance at NASA*. University of Chicago Press.

Vaughan, Diane. 1999. “The Dark Side of Organizations: Mistake, Misconduct, and Disaster.” *Annual Review of Sociology* 25: 271–305. https://doi.org/10.1146/annurev.soc.25.1.271.

Warr, Mark. 2002. *Companions in Crime: The Social Aspects of Criminal Conduct*. Cambridge University Press.

Warren, Danielle E. 2003. “Constructive and Destructive Deviance in Organizations.” *Academy of Management Review* 28 (4): 622–32. https://doi.org/10.5465/amr.2003.10899440.

Weick, Karl E., and Kathleen M. Sutcliffe. 2007. *Managing the Unexpected: Resilient Performance in an Age of Uncertainty*. 2nd ed. Jossey-Bass.

Willison, Simon. 2026. “An AI Agent Published a Hit Piece on Me [Blog Post.” Newsletter. *Simon Willison’s Weblog*, February 12. https://simonwillison.net/2026/Feb/12/an-ai-agent-published-a-hit-piece-on-me/.

Wong, Matteo. 2026. “OpenAI Has Gone Rogue: It’s Impossible to Know the Extent of the AI-Hacking Crisis.” *The Atlantic*, September 28. https://www.theatlantic.com/technology/2026/09/ai-hacks-infestation/688806/.

Wortley, Richard. 2001. “A Classification of Techniques for Controlling Situational Precipitators of Crime.” *Security Journal* 14 (4): 63–82. https://doi.org/10.1057/palgrave.sj.8340098.

Yar, Majid. 2005. “The Novelty of ‘Cybercrime’: An Assessment in Light of Routine Activity Theory.” *European Journal of Criminology* 2 (4): 407–27. https://doi.org/10.1177/147737080556056.